\documentclass[12pt]{article}
\usepackage[utf8]{inputenc}
\usepackage{amsmath}
\usepackage{bbm}
\usepackage{setspace}
\usepackage{subcaption}
\usepackage{graphicx}
\usepackage{appendix}
\usepackage{authblk}
\usepackage{multirow}
\usepackage[round]{natbib}
\usepackage{color}

\usepackage[left=2.5cm,right=2.5cm,top=2.5cm,bottom=2.5cm]{geometry}

\newtheorem{Lemma}{Lemma} 
\newtheorem{Prop}{Proposition}
\newcommand{\Esp} {\mathbbm{E}}
\newcommand{\Corr}{\mathsf{Corr}}
\newcommand{\Cov}{\mathsf{Cov}}
\newcommand{\Var}{\mathsf{Var}}
\newcommand{\R}{ \mathbbm{R} }
\newcommand{\Ind}{\mathbbm{1}}
\newcommand{\Pro}{\mathbbm{P}}

\providecommand{\keywords}[1]
{
  \small	
  \textit{Key words:} #1
}

\title{Beta-CoRM: A Bayesian Approach for $n$-gram Profiles Analysis}
\author{José A. Perusquía$^{1,2}$, Jim E. Griffin$^3$, Cristiano Villa$^{4,5}$}
\affil{\small{Department of Mathematics, Faculty of Sciences, UNAM,Mexico$^1$\\
Department of Probability and Statistics, IIMAS, UNAM, Mexico$^2$
\\ Department of Statistical Science, University College London, UK$^3$
\\
Division of Natural and Applied Sciences, Duke Kunshan University, China$^4$
\\
School of Mathematics, Statistics and Physics, Newcastle University, UK$^5$}}
\date{}

\begin{document}
\setstretch{1.2}

\maketitle
\begin{abstract}

$n$-gram profiles have been successfully and widely used to analyse long sequences of potentially differing lengths for clustering or classification. Mainly, machine learning algorithms have been used for this purpose but, despite their predictive performance, these methods cannot discover hidden structures or provide a full probabilistic representation of the data. A novel class of Bayesian generative models designed for $n$-gram profiles used as binary attributes have been designed to address this. The flexibility of the proposed modelling allows to consider a straightforward approach to feature selection in the generative model. Furthermore, a slice sampling algorithm is derived for a fast inferential procedure, which is applied to synthetic and real data scenarios and shows that feature selection can improve classification accuracy.

\end{abstract}


\section{Introduction}
\label{Intro}
Classification, as the statistical problem of identifying the group to which an observation belongs, is at the core of many fields of research. From a practical point of view, the statistical methods and algorithms used for this purpose depend, among other things, on the type of data, the number of groups (if known \textit{a priori}) and whether we have prior knowledge on how the groups look, for example, if we have access to labeled data. Although the process of labelling can certainly be expensive and perhaps quite complicated, it provides the data with a rich structure that can be exploited to efficiently learn the underlying generative process of the data and provide, among other things, accurate classification algorithms. In this direction, for certain applications, where observations are characterised by a set of binary attributes some of the most widely used models belong to the class of machine learning algorithms, such as (boosted) decisions trees, random forests and support vector machine, among others.

The algorithms just mentioned, and in particular the boosted methods, have become quite popular because of their superb predictive performance; however, one major drawback of these algorithms is that they can not be used for other tasks, such as finding unobserved structure in the data or providing a probabilistic representation of uncertainty for which Bayesian inference is particularly attractive. That is why in this paper we centre our attention on a novel class of Bayesian supervised learning models designed for observations characterised by a set of independent binary attributes. In particular we are interested in $n$-gram profiles, that have been used for a long time in many fields of research such as natural language processing \cite[see\textit{ e.g.}][]{Cavnar94}, cyber security \cite[see \textit{e.g.}][]{Kolter2004,Geng2011} and genomics \cite[see \textit{e.g.}][]{Tomovic2006}, to mention a few. 

Broadly speaking, an $n$-gram sequence is a contiguous sequence of $n$ elements, that depending on the field of interest, can be letters, numbers, words, computer commands, etc. These are used to summarise observations, such as documents or computer programs. From a theoretical point of view, summarising by $n$-grams sequences has the advantage that more information is retained compared to using single elements as attributes. This is something that has been widely exploited in natural language processing applications, such as language detection, spelling correction or speech recognition, where one of the main objectives is to predict the upcoming word given the previous $n$ words under Markovian assumptions. Prediction of new elements is not the only use of $n$-grams and their presence/absence or their frequencies can also be used as features to characterise observations. These features are often used for clustering and classification tasks such as anomaly detection. 

Although $n$-grams have yielded good results in many fields of research it is important to acknowledge that there are still some theoretical and computational drawbacks that need to be kept in mind when using them. For instance, for certain applications such as cyber security, the biggest computational challenge is the curse of dimensionality since even for small data sets the number of $n$-grams can easily reach the order of hundreds of thousands, or even billions \cite[see \textit{e.g.}][]{raff}, and hence, making it impossible to use all $n$-grams in an analysis. That is why a feature selection procedure is required as a first step. Deciding how many and which $n$-grams to use is vital and has lead to an interesting and active area of research. Within the computer science community some of the most common criteria are either frequency-based, e.g. using the features that appear at least $K$ times as in \cite{Geng2011} or information-based, e.g. using the $M$ most important features with respect the mutual information gain as in \cite{Kolter2004}. 
When there is a huge number of initial $n$-grams to be considered, this thinning process can often be so extreme (reducing by several orders of magnitude) that it has the practical effect of diminishing the inherent dependence of the remaining $n$-grams to such a degree that they can be considered close to independent.

In this paper, we take the large number of remaining $n$-grams as motivation for a Bayesian nonparametric approach. The Bayesian approach to statistics can be motivated by the notion of exchangeability; however, in many real-life situations, for example, where there is a need to model hierarchical data, a more general dependence structure is required. Within the Bayesian nonparametric framework, several approaches have been proposed to model data that is assumed to be generated from different, although related, populations \citep{camerlenghi2019}. For example, one could work with nested processes, such as the nested Dirichlet process (and its variations) \cite[see \textit{e.g.}][]{rodriguez2008nDP,camerlenghi2019}, hierarchical normalised random measures \citep{camerlenghi} which include the well-known hierarchical Dirichlet process \citep{Teh_hdp}, and additive processes \citep{Griffin13, LNP14}.

Working with hierarchies of discrete nonparametric priors has become quite popular since it allows the so-called effect of sharing of information. In this paper, we centre our attention on the class of compound random measures \citep{jim+fabrizio,jim+fabrizio2} as the discrete nonparametric prior. We believe that compound random measures, are particularly attractive for $n$-gram profiles since they allow us to build correlated measures that characterise and differentiate the groups and which yield a flexible model whose complexity grows with the size of the data. In particular, the beta-CoRM approach proposed here is a flexible probabilistic model, which allows us to define a hierarchy across groups and (contrary to other hierarchies of discrete nonparameteric priors) at the same time allows us to differentiate the importance of the shared features across groups. This is especially useful for finding the most influential features in each group from a common list of features. Furthermore, the structure of our model allows us to introduce a generalisation that can be used for the vital task of feature selection in the high-dimensional setting of $n$-grams, and hence, overcome the curse of dimensionality. 

The remainder of the paper is as follows: in Section~\ref{ngramProf} we provide a more detailed description of $n$-gram profiles and their applications; in Section~\ref{betaCoRM} and Section~\ref{featSel} we present our proposed model and the generalisation to allow a feature selection step; in Section~\ref{Est} we detail the posterior inference of the model where a slice sampling technique is developed for a faster update mechanism; in Section~\ref{Sensitivity} we present a prior sensitivity analysis paying special attention to the generalised version; in Section~\ref{Examples} we present the results of applying our proposed methodologies to real-case applications found in the cyber security realm; lastly in Section~\ref{conclusion} we provide some general conclusions and future work.

\section{$n$-gram Profiles and Discrimination among Groups}
\label{ngramProf}
Historically, $n$-grams have played a key role in many applications related to natural language processing \citep{Suen79} since they tend to be robust against spelling errors, they contain more information than using one element by itself and are relatively simple to obtain and understand \cite[][]{Tomovic2006,Geng2011}. Although $n$-gram models have yielded good results in most of the areas of research where they have been used, they also represent a computational challenge because in certain applications the number of $n$-grams explodes. This is something that has been seen especially in cyber security applications related to malware detection and classification where the number of $n$-grams reach the order of millions and even billions \cite[see \textit{e.g.}][]{Kolter2004,raff}. Therefore, there is a need to first decide which $n$-grams to use which can be done through frequency-based, information-based or more complex feature selection methods.

Within the natural language processing framework, an $n$-gram is usually defined as a contiguous sequence of $n$ items from a given text. Depending on the application these items can be words, letters, numbers, symbols, etc. For instance, if we consider the example: \textit{cats are connoisseurs of comfort} and use words as the items, then the set of 1-grams is $\{$cats, are, connoisseurs, of, comfort$\}$; the set of 2-grams is $\{$cats are, are connoisseurs, connoisseurs of, of comfort$\}$, the set of 3-grams is
$\{$cats are connoisseurs, are connoisseurs of, connoisseurs of comfort$\}$, and the set of 4-grams is
$\{$cats are connoisseurs of, are connoisseurs of comfort$\}$. We can then appreciate that the length of the set of $n$-grams depends on the value of $n$ and the length of the text considered. In the previous example, we have a sentence comprised of five words so that we have five 1-grams, four 2-grams, three 3-grams and two 4-grams. 

In this way, if the $i$-th observation is a text of $K_i$ items, there exist at most $K_i-n+1$ different $n$-grams (since there is always the possibility of having duplicates). Hence, for data comprised of $N$ observations we are then able to extract a vocabulary of at most $\sum_{i=1}^{N}(K_i-n+1)$ unique $n$-grams, which are used as features to create profiles that completely identify groups and observations within them. In consequence, being able to accurately model these profiles is vital to understand the structure of the data for replication purposes and for classification of new observations. In practice two main approaches for this modelling have been considered: 1) modelling the number of times each  $n$-grams appears or 2) reducing the data to the presence or absence of each $n$-gram, that is, binary features. In this paper, we will consider the latter which is natural in the application to cyber security that we explore. 

Within the computer science realm, the use of $n$-grams as binary features has been considered for detecting masquerade attacks using UNIX commands \cite[\textit{see e.g.}][]{Geng2011}, for detecting and classifying malicious software \cite[\textit{see e.g.}][]{Kolter2004,Pektas2011} using the binary content of the malware and for characterising the behaviour of a program using the kernel calls in order to detect anomalies \cite[\textit{see e.g.}][]{Marceau}. From a Bayesian perspective, these applications just as other cyber security tasks represent a relatively new, interesting and challenging area of research since most of the methods used to detect the attacks belong to the class of machine learning supervised models, such as (boosted) decision trees. For the application considered in this paper we centre our attention on the malware classification problem for which we are able to straightforward apply and see the advantages of the beta-CoRM models in particular of the generalised versions that allows us to perform a feature selection process to select an optimal set of $n$-grams and reduce the uncertainty of the data.

More precisely, our interest mainly centres on analysing grouped data, that is, data whose observations are already categorised into $d$-groups (e.g. the different types of malware such as worms, spyware, adware, ransomware, etc.). Then the methodology consists on extracting a set of $M$ different $n$-grams that are assumed to completely characterise the groups and the observations through their presence/absence. Since these are binary attributes we can think of the data as a grouped binary matrix, that is, a binary matrix containing $d$ sub matrices with the $j$-th group having $n_j$ observations so that the total number of rows is $\sum_j n_j=N$ and the total number of columns is given by $M$. Then, if we denote  this grouped binary matrix by $X$, we can define its entries as $X_{kji}\in\{0,1\}$ where $i\in\{1,...,M\}$, $j\in\{1,...,d\}$ and $k\in\{1,...,n_j\}$, that is, the rows of $X$ are indexed by the pair $(j,k)$ which respectively indicate the group and observation within that group so that $X_{kji}=1$ indicates that the $k$-th observation of the $j$-th group contains the $i$-th $n$-gram and $X_{kji}=0$ otherwise. We will write $X_{kj} = (X_{kj1}, \dots, X_{kjM})$.
\color{black}

The grouped binary matrices play a vital role in the Bayesian models we present in the next two sections. As we thoroughly describe in the upcoming sections, the models considered allows us to provide an individual probabilistic characterisation of the importance of each $n$-gram at a group level which is used to differentiate groups. Furthermore, our modelling framework provides us with the tools to perform feature selection to find an optimal set of $n$-grams which provide the largest differences between the groups and hence, maximise discrimination among them.

\section{Beta-CoRM}
\label{betaCoRM}
In this section, we present a novel Bayesian approach to supervised learning that builds on a special type of $d$-dimensional vector of completely random measures (CRM's) \citep{kingman1}, known as \textit{compound random measures} (CoRM's) \citep{jim+fabrizio}. Since their introduction in \cite{kingman1}, CRM's have become essential for most Bayesian nonparametric models. Theoretically, completely random measures, can be viewed as the sum of three components: a deterministic part, a term with a finite number of fixed points of discontinuity with random jump heights and a term with random points of discontinuity and random jump heights. Among practitioners it is quite common to consider completely random measures without fixed points of discontinuity so that, if $\Psi$ is a CRM, then it can be represented as,
\begin{eqnarray}\label{CRM}
\Psi&=&\sum_{i=1}^{\infty}J_{i}\delta_{\xi_i},
\end{eqnarray}
where both the locations of the jumps, $\xi_i$'s, and the height of the jumps, $J_i$'s, are random and where $\delta_x$ represents the Dirac delta measure which places mass one in $x$. Moreover, $\Psi$ is completely characterised by 
\begin{eqnarray*}
    \Esp\left(\exp(-u\Psi(A)\right))&=&\exp\left(-\int_{\R^+}(1-\exp(-us))\nu(A,ds)\right),
\end{eqnarray*}
where $\nu$ is referred to as the L\'evy measure and contains all the information about the distribution of $\xi_i$'s and the $J_i$'s \citep{kingman1}. For the purposes of this paper, we focus on the homogeneous case, so that
\begin{eqnarray*}
    \nu(dx,ds)&=&\rho(ds)\alpha(dx),
\end{eqnarray*}
where $\rho$ is a measure on $\R^+$ and $\alpha$ is a non-atomic measure. Therefore, the locations of the jumps are independent of their heights. 

In this paper we shall centre our attention on the beta process, which is completely characterised by its homogeneous L\'evy measure
\begin{eqnarray}
\label{levyBP}
\nu(d\omega,dp)&=&c(\omega)p^{-1}(1-p)^{c(\omega)-1}dpB_0(d\omega),
\end{eqnarray}
where $c(\omega)$ is a concentration function and $B_0$ is a finite fixed measure on $\Omega$. This L\'evy measure can then be interpreted as a measure on $\Omega\times[0,1]$, hence, one can see that the beta process concentrates its jumps in the interval $[0,1]$, so that they can be used as the parameters of an infinite sequence of Bernoulli random variables. As described in \cite{thibaux+jordan}, it is common among practitioners to consider $c(\omega)=c$, so that $c$ is a concentration parameter. As for $B_0$, this measure can be continuous, discrete or a mix of both types. In the Bayesian nonparametric literature the most common choice is absolutely continuous. This choice is particularly useful for factorial models like the Indian buffet process (and related models) where the interest centres on inference about the infinite number of unknown latent factors  \citep[see {\it e.g.}][]{griffiths+zoubin, griffiths+zoubin2, thibaux+jordan}.

Completely random measures are appealing structures due to their almost sure discreteness, which means that their realisations are discrete with probability one. This characteristic allows us to use CRM's to model data generated by a discrete distribution or to use them as the basic building block in mixture models. With this in mind, a compound random measure defines $d$ correlated measures by perturbing the jumps of a completely random measure, for example, considering a purely atomic completely random measure with a directing homogeneous L\'evy measure $\nu$ as in \eqref{CRM}, and if we let $\mu_j$ represent the $j$-th random measure then,
\begin{eqnarray}
\mu_j&=&\sum_{i=1}^{\infty}m_{ji}J_i\delta_{\xi_i},\hspace{2cm}m_{1i},...,m_{di}\stackrel{\text{iid}}{\sim}h,
\end{eqnarray}
where the $m_{ji}$’s are the perturbation coefficients, which are called scores, that identify specific features of the j-th random measure and $h$ is probability distribution, which is called the score distribution. 

Therefore, CoRM's are completely characterised through the distribution $h$ and the directing L\'evy measure $\nu$ of the completely random measure. This can be seen by the fact that the directing L\'evy measure of $\mu_j$ is
\begin{eqnarray*}
    \nu_j(dx,ds)&=&\rho_j(ds)\alpha(dx),
\end{eqnarray*}
where for each marginal process we have that,
\begin{eqnarray*}
    \rho_j(ds)&=&\int z^{-1}h(s/z)ds \nu(dz).
\end{eqnarray*}
From this integral representation, it is interesting to notice that compound random measure can be defined by specifying the score distribution and either the desired marginal processes, $\mu_j$, or the directing completely random measure. An interesting example occurs when each marginal process is a beta process with directing L\'evy measure as in \eqref{levyBP} and a beta$(a,1)$ score distribution, then the directing completely random measure can be seen as the sum of a beta process with a compound Poisson process (if $c>1$). 
For a complete theoretical study on CoRM's and some of their applications the reader can refer to \citeauthor{jim+fabrizio} (\citeyear{jim+fabrizio}, \citeyear{jim+fabrizio2}).

\subsection{Construction}
From the broad description given above, compound random measures are particularly attractive for grouped data and hence, for supervised learning. Since our interest centres on the probabilistic modelling of matrices with binary entries,  a suitable approach is to consider a beta-Bernoulli model. That is, we assume that the entries of the matrix are Bernoulli random variables with unknown parameter $\theta$ which we assume it follows a beta distribution. Within the Bayesian parametric framework, this choice yields a conjugate model so that fast inferential procedures can be carried out. In a Bayesian nonparametric setting this can be achieved by choosing a beta process (BP) $B$, as the directing completely random measure defined on a suitable space of features, henceforth denoted as $\Omega$.

For an $n$-gram profiles analysis a factorial model could also be used to discover some latent structure; however, in this paper we are interested in modelling the data directly. In order to do so, a discrete base measure $B_0$, on the set of $n$-grams ($\Omega$) might be a more suitable choice. Hence, 
\begin{eqnarray}\label{B0}
B_0&=&\sum_{i=1}^{\infty}q_i\delta_{\omega_i},
\end{eqnarray}
where the set of jumps, $q_i$'s, are random variables defined in the unit interval, and the set of locations $\omega_i$'s, work as deterministic labels to distinguish the different $n$-grams which are independent by construction, and hence, suitable for the applications we consider in this paper. This is a particularly interesting approach since the beta process will share the same atoms as $B_0$ with corresponding jumps $p_i$ sampled from a beta distribution $(cq_i,c(1-q_i))$ \citep{thibaux+jordan}. Therefore, $B$ has the following discrete representation 
\begin{eqnarray}\label{BP}
B&=&\sum_{i=1}^{\infty}p_i\delta_{\omega_i}.
\end{eqnarray}

By doing so, the set of jumps $p_i$ can be thought as the probability that an observation regardless of the class has the corresponding $n$-gram, and for each of the $d$ correlated groups these weights are perturbed by the scores $m_{ji}$. The perturbed coefficients $m_{ji}p_i$ represent the probability of observing each $n$-gram for each of the $d$ groups; choosing a score distribution on $(0, 1)$ ensures that these probabilities are properly defined. For the purposes of this paper we restrict our attention to a beta score distribution and in particular, to a beta distribution with shape parameters $(a,1)$, so that the expectation is $a/(a+1)$. Then as we shall see in Section~\ref{featSel} this choice represents a compelling approach to feature selection when working with a generative model. In this way, for the $j$-th group we have a marginal process given by
\begin{align}\label{Bj}
B_j&=\sum_{i=1}^{\infty} m_{ji}p_i \delta_{\omega_i}&m_{ji}\stackrel{\text{ind}}{\sim}\text{beta}(a,1).
\end{align}

Finally, the generative process is fully described by assuming that each observation $X_{kj}$ in group $j$ of the binary matrix $X$, follows a Bernoulli process with corresponding base measure $B_j$, so that for $k=1,\ldots,n_j$,
\begin{align}\label{Xkj}
X_{kj}&=\sum_{i=1}^{\infty} x_{kji}\delta_{\omega_i}&x_{kji}\sim\text{Ber}(m_{ji}p_i).
\end{align}

We believe that this Bayesian model provides an interesting approach to supervised learning with $n$-gram profiles. The structure of compound random measures allows us to represent the importance of each feature for each of the classes through the perturbation coefficients. This is particularly useful in situations where we expect to differentiate classes through differences found in some of the features.

\subsection{Properties}
\label{properties}
Now that the model has been described it is important to analyse its properties to fully understand the generative process and the role of the hyperparameters in the learning process.  For the directing discrete beta process we can obtain both the expectation and the variance, this will provide insights into the role of the the jumps $q_i$ and the concentration parameter $c$.
\begin{Prop}
Let B be a beta process with discrete base measure as in \eqref{BP} and \eqref{B0} respectively. Then
\begin{enumerate}
\item $\Esp(B)=B_0$
\item $\mathsf{Var}(B)=\frac{1}{c+1}\sum_{i=1}^{\infty}q_i(1-q_i)$.
\end{enumerate}
\end{Prop}

The proof of Proposition 1 is straightforward and can be found in \ref{proofs}. The properties stated in Proposition 1 provide important information about the parameters of the directing beta process and the role they have in the generative process. For instance, the $q_i$'s represents our prior knowledge on the global probabilities, and the concentration parameter $c$, controls the similarity between the $p_i$'s and the $q_i$'s. Then as $c\rightarrow\infty$, we have that the $\mathsf{Var}(p_i)\rightarrow0$ and hence, $p_i\stackrel{\text{a.s.}}{\rightarrow}q_i$, therefore, large values of $c$ should be used when there is a strong prior belief that the $q_i$'s are good estimates of the $p_i$'s. On the other hand, for values of $c$ close to 0 then $p_i$ will be either close to 1 or 0 with probabilities $q_i$ and $(1-q_i)$ respectively.

For the correlated measures $B_j$'s, that characterise each group, useful properties can also be derived, like their expectation and their variance (at a fixed point of discontinuity) as stated in the following proposition.
\begin{Prop}
    Let $B$ be a beta process defined as in Proposition 1 and $B_j$ denote the $j$-th measure defined as in \eqref{Bj}, then
    \begin{enumerate}
\item $\Esp(B_j|B)=\frac{a}{a+1}B$ and hence $\Esp(B_j)=\frac{a}{a+1}B_0$.
\item For a fixed feature $\omega_i$ 
\begin{eqnarray*}
\Var(B_j(d\omega_i))&=&\left(\frac{aq_i}{a+2}\right)\left(\frac{(1-q_i)(a+1)^2+q_i(c+1)}{(c+1)(a+1)^2}\right).
\end{eqnarray*}
\end{enumerate}
\end{Prop}

We can then immediately appreciate that larger values of the shape parameter $a$ will make the correlated measures closer to the discrete base measure $B_0$. Whereas smaller values will yield correlated measures with jumps closer to zero. Moreover, since the shared jump $p_i$ introduces dependence between the jump heights in each measure, the covariance and the correlation at each location $\omega_i$ can also be obtained as stated in the following proposition.

\begin{Prop}
Let $B$ be a beta process defined as in Proposition 1 and $B_j$ and $B_k$ denote the $j$-th and the $k$-th measure defined as in \eqref{Bj}, then for a fixed feature $\omega_i$ 
\begin{eqnarray*}
    \Cov(B_j(d\omega_i),B_k(d\omega_i))&=&\left(\frac{a}{a+1}\right)^2\frac{q_i(1-q_i)}{c+1}
\end{eqnarray*} 
and 
\begin{eqnarray*}
    \Corr(B_j(d\omega_i),B_k(d\omega_i))=\frac{a(a+2)(1-q_i)}{(a+1)^2(1-q_i)+q_i(c+1)}.
\end{eqnarray*}
\end{Prop}

From these properties we can immediately obtain the probability of an observation having the feature $\omega_i$, which does not depend on $c$ and is given by
\begin{equation*}
\Pro(x_{kji}=1)=\Esp(B_j(d\omega_i))=\frac{a}{a+1}q_i.
\end{equation*}
Then the joint distribution can be further generalised to consider all groups by simply obtaining the $d$-th moment of a beta distribution with parameters $(cq_i,c(1-q_i))$, yielding 
\begin{eqnarray*}
\Pro\left(\prod_{j=1}^dx_{kji}=1\right)=\Esp\left(\prod_{j=1}^dB_j(d\omega_i)\right)=\left(\frac{a}{a+1}\right)^d\left(\prod_{j=0}^{d-1}\frac{cq_i+j}{c+j}\right).
\end{eqnarray*}

Finally, it is also interesting to notice that the covariance is the difference between the joint probability of two observations in different groups having the feature $\omega_i$ and the distribution assuming independence, that is, for the $n$-th and $m$-th observation in the $j$-th and $k$-th group respectively,
\begin{eqnarray*}
\Cov(B_j(d\omega_i),B_k(d\omega_i))
&=&\left(\frac{a}{a+1}\right)^2\left(\frac{cq_i^2+q_i}{c+1}\right)-\left(\frac{a}{a+1}\right)^2q_i^2.
\end{eqnarray*}

\section{Feature Selection}
\label{featSel}

Now that the discrete beta-CoRM approach for grouped binary matrices has been fully described, in this section we present a generalisation of the model that  yields a natural feature selection procedure which will be particularly useful if the feature space is a high-dimensional object and our goal is discrimination about $d$ groups. This approach arises naturally by noting that the density of the beta($a$,1) score distribution can be written as
\begin{eqnarray*}
    ax^{a-1}=x_0^a\frac{ax^{a-1}}{x_0^a}\Ind_{(0,x_0)}^{(x)}+(1-x_0^a)\frac{ax^{a-1}}{1-x_0^a}\Ind_{(x_0,1)}^{(x)}=(1-w)f(x)+wg(x)
\end{eqnarray*}
where $f(x)$ and $g(x)$ are truncated beta distributions on $(0,x_0)$ and $(x_0,1)$, respectively. For small $x_0$, this representation mimics the form of a spike-and-slab prior \citep[see {\it e.g.}][]{Mitchell, GeMc97, ishwaran2005} with $w$ the probability of ``including'' a variable and $g(x)$ the slab distribution. As for $g(x)$ its cumulative distribution function (cdf) is
\begin{eqnarray*}
G(x)&=&\frac{x^a-x_0^a}{1-x_0^a},
\end{eqnarray*} 
and 
\begin{eqnarray*}
\lim_{a\downarrow0}G(x)&=&\frac{\log(x_0)-\log(x)}{\log(x_0)},
\end{eqnarray*} 
with the corresponding probability density function (pdf)
\begin{eqnarray*}
g(x)&=&\frac{1}{\log(1/x_0)}\frac{1}{x},\hspace{1cm}x>x_0.
\end{eqnarray*}

\begin{figure}[ht]
    \centering
    \includegraphics[scale=.5]{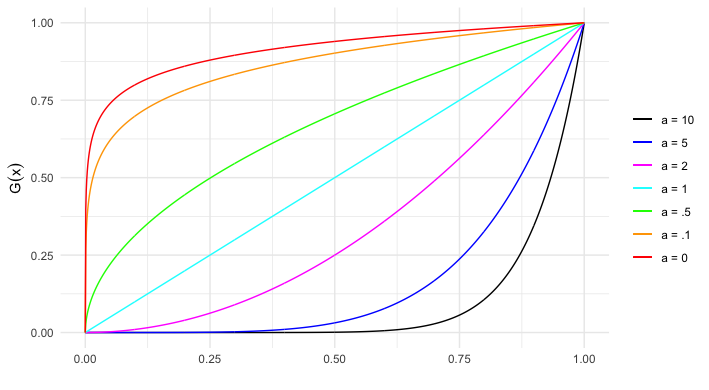}
    \caption{Cumulative distribution function of the slab distribution for different values of the score parameter $a$ and $x_0=.00001$.}
    \label{fig:slab}
\end{figure}

Therefore, we can understand the prior distribution as a spike-and-slab prior where $a$ controls the size of the spike with a small value of $a$ implying a small value of $w$. If $a$ is close to zero, the pdf of the slab is approximately $g(x)$. To have a better understanding of the score parameter $a$ in the beta-CoRM model, in Figure~\ref{fig:slab} we present a graphical representation of the cdf of the slab distribution for different values of $a$ and with $x_0=.00001$. Now, since the beta$(a,1)$ random variables moderate $p_i$, it can be immediately appreciated that for small $a$ the prior expects some of the scores $m_{ji}$ to be close to zero and hence the respective products $m_{ji}p_i$ to be close to zero as well and with $w$ controlling the proportion close to ``zero''.

With this spike-and-slab interpretation of the beta distribution in mind, we can then provide a natural way to generalise the beta-CoRM model to allow for  a feature selection procedure by giving a unique score parameter $a_i$ to each variable $\omega_i$ leading to
\begin{align*}
    p_i&\sim \text{beta}(cq_i,c(1-q_i)),&\\
    m_{ji}&\sim\text{beta}(a_i,1),&j\in{1,\ldots,d}\\
    \intertext{and}
    x_{kji}&\sim\text{Ber}(m_{ji}p_i),&k\in{1,\ldots,n_j},
\end{align*}
with all the $a_i$'s having a common prior distribution. Then the posterior estimates of these score parameters can be used to determine which are the best discriminative features (such as $n$-grams) by noting that features with a ``small'' score parameter should be preferred since they yield more distinguishable groups which is particularly useful for supervised classification. As we shall see in Section 6, this adjustment can lead to better classification performance.

\section{Posterior Inference}
\label{Est}

In order to derive the full conditional distributions, we first restrict our attention to the beta-CoRM model described in Section~\ref{betaCoRM}, where a single score parameter $a$ is used. Then we shall see that a straightforward generalisation is achieved when using a set of score parameters. 

\subsection{Beta-CoRM}
 Due to the discrete nature of the model, the joint posterior distribution of the correlated random measures $\{B_j\}_j$ and the directing beta process $B$ is the product of the posterior distribution of the random variables associated to each atom, that is, $p_i$ and the set of scores $m_{1i},...,m_{di}$. Therefore, we can analyse the posterior density on each atom. So, if we consider the $i$-th feature, the posterior density (up to proportionality) is given by
\begin{eqnarray}
\label{post}
&&\left(\prod_{j=1}^d(p_im_{ji})^{x_{\cdot ji}}(1-p_im_{ji})^{n_j-x_{\cdot ji}}\right)\left(p_i^{cq_i-1}(1-p_i)^{c(1-q_i)-1}\right)\prod_{j=1}^dm_{ji}^{a-1}\nonumber\\
&=&\left(p_i^{cq_i+x_{\cdot \cdot i}-1}(1-p_i)^{c(1-q_i)-1}\right)\prod_{j=1}^dm_{ji}^{x_{\cdot ji}+a-1}(1-p_im_{ji})^{n_j-x_{\cdot ji}},
\end{eqnarray}
where $x_{\cdot ji}=\sum_{k=1}^{n_j}x_{kji}$ and $x_{\cdot\cdot i}=\sum_{j=1}^d x_{\cdot ji}$. From \eqref{post} we can immediately appreciate that due to the presence of the $d$ terms $(1-m_{ji}p_i)$, the joint and the conditional distributions do not have a known form. Therefore, a Gibbs sampling algorithm cannot be applied directly to this posterior distribution. One way to address this issue is with the introduction of a set of latent variables $\{u_{kji}\}$ that allows us to define an artificial measure $B_{kj}$ as the base measure for the $k$-th observation in the $j$-th group, that is,
\begin{align}
\label{slice}
B_{kj}&=\sum_i y_{kji}p_i \delta_{\omega_i},&y_{kji}=\Ind(u_{kji}<m_{ji})\sim\text{Ber}(m_{ji})\nonumber\\
\intertext{and}
X_{kj}&=\sum_i x_{kji}\delta_{\omega_i},&x_{kji}\sim\text{Ber}(y_{kji}p_i).
\end{align}

This approach is based on the idea of slice sampling \citep{slice_samp}, where a set of latent variables that preserve the marginal distribution are introduced. Slice sampling schemes have become widely used in Bayesian nonparametric models since they yield efficient computational methods for the infinite dimensional objects that are at their core. The reader can refer to \cite{griffin_walker} and the references therein for an overview of the computational issues found in some nonparametric models and the approaches used to address them. As for the normalised completely random measures, the slice sampling technique is useful in order to introduce a random truncation point and hence, consider only a random finite number of jumps. In our case the slice sampling approach that we propose allows us to create these infinite activity measures $\{B_{kj}\}_{k,j}$, that yield a suitable augmented likelihood from which we can recover the original one by integrating out the latent variables as stated in the following Lemma.
\begin{Lemma}
The discrete beta-CoRM defined by equations \eqref{BP}, \eqref{Bj} and \eqref{Xkj} is equivalent to the augmented model in \eqref{slice}.
\end{Lemma}

In order to prove Lemma 1 it is sufficient to note that the likelihood for a specific observation $x_{kji}$ is a mixture of a degenerate distribution and a Bernoulli distribution with parameter $p_i$ with respective weights $(1-m_{ji})$ and $m_{ji}$. The complete details can be found in the \ref{proofs}. Now, with this form of the likelihood, the complete augmented posterior (up to proportionality) is given by
\begin{eqnarray}
\label{augmented_post}
&&\left(p_i^{cq_i-1}(1-p_i)^{c(1-q_i)-1}\right)\left(\prod_{j=1}^d\prod_{k=1}^{n_j}\left(\delta_0^{x_{kji}}\right)^{(1-y_{kji})}p_i^{x_{kji}y_{kji}}(1-p_i)^{(1-x_{kji})y_{kji}}\right)\nonumber\\
&&\times\left(\prod_{j=1}^d\left[m_{ji}^{a-1}\prod_{k=1}^{n_j}m_{ji}^{y_{kji}}(1-m_{ji})^{1-y_{kji}}\right]\right).
\end{eqnarray}

From \eqref{augmented_post} we can immediately notice that the conditional posterior distributions are
\begin{align*}
p_i|\mathbf{X},\mathbf{Y}&\sim\text{beta}\left(\sum_{j=1}^d\sum_{k=1}^{n_j}x_{kji}y_{kji}+cq_i,\sum_{j=1}^d\sum_{k=1}^{n_j}(1-x_{kji})y_{kji}+c(1-q_i)\right),
\end{align*}
\begin{align*}
m_{ji}|\mathbf{Y}&\sim\text{beta}\left(a+\sum_{k=1}^{n_j}y_{kji},1+n_j-\sum_{k=1}^{n_j}y_{kji}\right),&j\in\{1,...,d\},
\end{align*}
and
\begin{align*}
y_{kji}|\mathbf{X},m_{ji},p_i&\sim\begin{cases}
\delta_{1}&\text{if} \hspace{.5cm}x_{kji}=1\\
\text{Ber}\left(\frac{(1-p_i)m_{ji}}{1-p_im_{ji}}\right)&\text{if}\hspace{.5cm} x_{kji}=0
\end{cases}&k\in\{1,...,n_j\}.
\end{align*}
Hence, a straightforward Gibbs sampling algorithm can be used for the posterior inference. Clearly, it is also important to notice that this slice sampling technique is also valid when we use individual score parameters $a_i$ rather than a global parameter $a$. In this case, however, the posterior distribution of the scores $\{m_{ji}\}_{j=1}^d$ is a beta distribution with shape parameter $a_i+\sum_{k=1}^{n_j}y_{kji}$ and the rest remains the same. The full conditional distributions of the other parameters are as follows.

\subsubsection{Full conditional distribution of $c$}
\label{post_inf_c}
For the prior specification of the concentration parameter we can recall from Section~\ref{properties} that $c$ is a positive parameter that primarily modulates the variance of the weights of the beta process, so that when $c\rightarrow\infty$, $p_i\stackrel{\text{a.s.}}{=}q_i$ for all $i$. Then, using \eqref{augmented_post} we have that the full conditional density of $c$ is proportional to
\begin{eqnarray*}
    \label{post_c}
    f(c|\textbf{p},\textbf{q},\theta_c)&\propto&\left(\prod_{i=1}^M\frac{p_i^{cq_i-1}(1-p_i)^{c(1-q_i)-1}}{B(cq_i,c(1-q_i))}\right)f(c|\theta_c),\\
\end{eqnarray*}
where $f(c|\theta_c)$ is the prior distribution, $\theta_c$ the set of hyperparameters and $M$ the number of active features. Clearly, this full conditional distribution does not have a closed form and to sample from it we follow the adaptive random walk Metropolis Hasting \citep{metro_hast} on $\log (c)$ so that a suitable Gaussian distribution can be used as the proposal distribution.

\subsubsection{Full conditional distribution of $a$}
\label{post_inf_a}

For the full conditional distribution of the score parameter $a$ we first notice that the full conditional density is proportional to
\begin{eqnarray*}
    \label{post_a}
    f(a|\textbf{M},\theta_a)&\propto&a^{Md}\exp{\left[a\sum_{j=1}^d\sum_{i=1}^M\log{m_{ji}}\right]}f(a|\theta_a),
\end{eqnarray*}
where $f(a|\theta_a)$ is the prior distribution and $\theta_a$ the corresponding hyperparameters. In this case and contrary to the concentration parameter we can directly observe that if $a\sim\text{gamma}(\lambda,\alpha)$ then we obtain a conjugate model so that the full conditional distribution of $a$ is gamma with parameters $(\lambda+Md,\alpha-\sum_{j=1}^d\sum_{i=1}^M\log{m_{ji}})$. Of course, we acknowledge that other prior distributions could be used and a Metropolis Hasting step would be required to sample from the posterior. This trade-off between computational ease and more general priors is something that we need to keep in mind for certain applications where fast algorithms are required.

\subsection{Generalised Beta-CoRM}
For the posterior inference of the generalised beta-CoRM model there is just a need to derive the full conditional distribution of the individual score parameters $a_i$, since the rest remains the same.

\subsubsection{Full conditional distribution of $a_i$}
\label{post_inf_ai}
The full conditional distribution for the individual score parameters $a_i$ follows the same reasoning and steps as for the global parameter $a$ by noting that in this case the full conditional density of each $a_i$ is proportional to
\begin{eqnarray*}
    \label{post_ai}
    f(a_i|\textbf{M},\theta_{a_i})&\propto&a_i^{d}\exp{\left[a_i\sum_{j=1}^d\log{m_{ji}}\right]}f(a_i|\theta_{a_i}),
\end{eqnarray*}
where $f(a_i|\theta_{a_i})$ is the common distribution of the scores with hyperparameters $\theta_{a_i}$. We can then notice that 
a gamma$(\alpha,\beta)$ prior could be used
for computational ease so that the full conditional of each score is a gamma distribution with parameters $(\alpha+d,\beta-\sum_{j=1}^d\log{m_{ji}})$. 

In this case however, we are also interested in analysing the effects of different priors since there is an inherent interest on doing feature selection. In particular, we are interested on scale-mixture gamma priors that allow us to define local and global shrinkage priors while still having a conjugate posterior for the score parameters, such as the gamma-gamma prior introduced in 
\cite{Griffin_Brown}, that is,
\begin{eqnarray}
    \label{gamma-gamma-prior}
    a_i|\lambda,\alpha_i&\sim&\text{gamma}\left(\lambda,\alpha_i\right),\nonumber\\
    \alpha_i|\phi,\kappa&\sim&\text{gamma}\left(\phi,\kappa\right),
\end{eqnarray}
where $\kappa$ is a scale parameter, $\lambda$ controls the behaviour of the distribution close to zero and $\phi$ controls the tail of the distribution. More general scale-mixture gamma priors such as the double gamma prior \citep{double_gamma} or the triple gamma prior \citep{triple_gamma} could also be considered.

\section{Prior Sensitivity Analysis}
\label{Sensitivity}
\begin{figure}[ht]
    \centering
    \includegraphics[scale=.5]{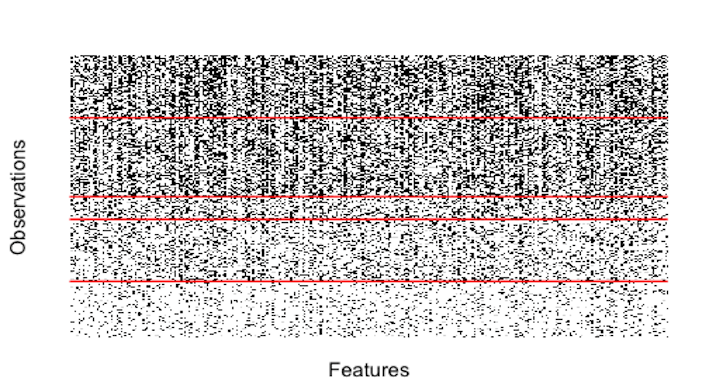}
    \caption{Synthetic data set composed of 5 imbalanced groups separated by the red lines with 250 total observations and 300 binary features.}
    \label{fig:synthetic}
\end{figure}

To analyse the effects of the prior on the beta-CoRM models proposed in the previous sections we consider a synthetic data set composed of 5 imbalanced groups with 250 total observations and 300 binary features as graphically represented in Figure~\ref{fig:synthetic}. We are interested in looking at the computational cost of posterior inference and the impact of feature selection on the predictive performance of the models. It is important to notice that for the predictive performance, the test set has been generated using the same parameters as the synthetic data. Furthermore, and for illustrative purposes on the feature selection process we obtain the optimal number of features by using the true labels of the test set. Of course, we acknowledge that on real-world applications this is not possible and hence, to find the ``best'' features we could rely on a cross validation procedure. 

\subsection{Prior Selection}
Since the interest mainly centres on the prior structure given to the score parameter(s) $a$ $(a_i's)$ we consider the same vague gamma prior on the concentration parameter $c$ for the different prior setups. As for the score parameters we can recall from Section~\ref{post_inf_a} that we are mainly interested in scale-mixture gamma priors and gamma vague priors since they allow us to have a conjugate structure on some of the parameters while introducing a shrinkage structure ideal for feature selection. In particular we centre our attention on vague gamma priors for $\lambda$ and $\alpha$ and on the gamma-gamma prior structure described in \eqref{gamma-gamma-prior} since it allows us to study quite interesting setups for different values of $(\lambda,\phi,\kappa)$ such as a half-Cauchy type prior ($\lambda=\phi=0.5$), the Lomax prior ($\lambda=1$) and the objective Lomax prior \citep{walker+villa} ($\lambda=\phi=\kappa=1$), which is a particular case of the Lomax distribution with shape and scale parameter equal to one. The objective Lomax prior is a particular type of objective distribution that has the property of not depending on the sampling distribution, but only on the parameter space, and to minimise an information criterion. Compared to other objective priors, it has the advantage of being proper, easing its implementation. 

To delve deeper into these particular cases, one can see that the marginal distribution of each $a_i$ is given by
\begin{eqnarray*}
    f(a_i)&=&\left(\frac{1}{\kappa}\right)^{\lambda}\frac{\Gamma(\lambda+\phi)}{\Gamma(\lambda)\Gamma(\phi)}a_i^{\lambda-1}\left(1+\frac{a_i}{\kappa}\right)^{-(\lambda+\phi)},
\end{eqnarray*}
so that the half-Cauchy type prior has density
\begin{eqnarray*}
    f(a_i)&=&\frac{\sqrt{\kappa}}{\pi\sqrt{a_i}}\left(\frac{1}{\kappa+a_i}\right),
\end{eqnarray*}
the Lomax prior has density
\begin{eqnarray*}
    f(a_i)&=&\left(\frac{\phi}{\kappa}\right)
    \left(1+\frac{a_i}{\kappa}\right)^{-(1+\phi)},
\end{eqnarray*}
and finally, the objective Lomax has density given by
\begin{eqnarray*}
    f(a_i)&=&(1+a_i)^{-2}.
\end{eqnarray*}

We can then appreciate from their distributions, that the half-Cauchy type prior has the heaviest tails and that the Lomax prior is the only one that can have a first finite moment whenever $\phi>1$.

With respect to the posterior inference, it is important to mention that for all these models and in order to save computational resources due to the large number of parameters sampled at each step, we consider an effective sample size of 1000 observations obtained from the posterior distribution using the MCMC scheme detailed in Section~\ref{Est} with 251,000 iterations, a burn-in of 1000 and a thinning of 250 to ensure low-correlated samples for the parameters of the beta-CoRM models. As of the computational time and despite the large number of simulations the four models required around 12 minutes to finish the posterior inference.

\subsection{Posterior Inference Analysis}
\label{post_inf_an}
With respect to the posterior inference analysis there are several aspects of the simulations results that we would like to contrast among the five prior structures considered. Firstly, in Table~\ref{post_inference_c} we present the median and the 95$\%$ credible interval for the only common parameter, that is, the concentration parameter $c$. It can be noted that quite consistent results are obtained with the only exception of the objective Lomax prior that yields values slightly shifted to the left. We believe this can be explained by the specific objective nature of this prior, where the minimization of information results in the prior being dominated by the data. Without this property, it is safe to assume that other prior distributions do not exhibit the same behavior.

\begin{table}[ht]
    \centering
    \begin{tabular}{lccc}
    \hline
    Model/Prior & L. Interval & Median & U. Interval\\
    \hline
    Beta-CoRM/Vague gamma & 4.8134  & 5.7047  & 6.8216 \\
    Gen. beta-CoRM/Vague gamma & 4.7233  & 5.7098  & 6.8284 \\
    Gen. beta-CoRM/Obj. Lomax  & 4.4747  & 5.4119  & 6.4652 \\
    Gen. beta-CoRM/Lomax & 4.7065  & 5.7183  & 6.8587 \\
    Gen. beta-CoRM/Half-Cauchy & 4.8882  & 5.8403  & 7.0668 \\
    \hline
    \end{tabular}
    \caption{Median and 95$\%$ credible interval comparison for the concentration parameter.}
    \label{post_inference_c}
\end{table}

Now we turn our attention to which might be the most interesting parameter for the beta-CoRM model which is the score parameter $a$ and the score parameters for the generalised versions. For this the first thing we would like to notice is the median and 95$\%$ credible interval in the beta-CoRM model displayed in Table~\ref{post_inference_a}.

\begin{table}[ht]
    \centering
    \begin{tabular}{lccc}
    \hline
    Model & Lower Interval & Median & Upper Interval\\
    \hline
    Beta-CoRM & 1.5317  & 1.6397  & 1.7519 \\
    \hline
    \end{tabular}
    \caption{Median and 95$\%$ credible interval for the score parameter}
    \label{post_inference_a}
\end{table}

\begin{figure}[ht!]
\begin{subfigure}{.5\textwidth}
\centering
  \includegraphics[scale=.3]{ 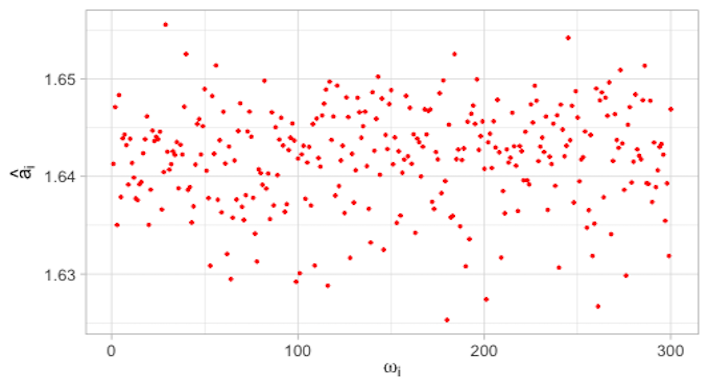}
  \caption{Vague gamma}
\end{subfigure}
\begin{subfigure}{.5\textwidth}
\centering
  \includegraphics[scale=.3]{ 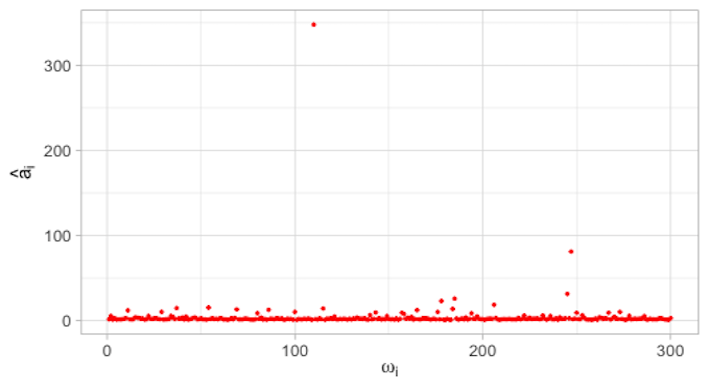}
  \caption{Objective Lomax}
\end{subfigure}
\begin{subfigure}{.5\textwidth}
\centering

  \includegraphics[scale=.3]{ 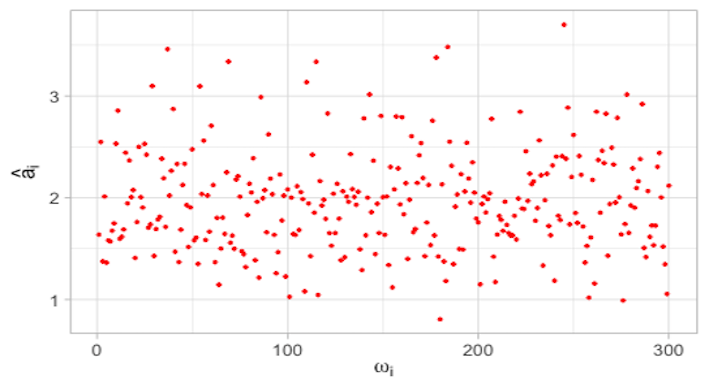}
  \caption{Lomax}
\end{subfigure}
    \begin{subfigure}{.5\textwidth}
\centering
  \includegraphics[scale=.3]{ 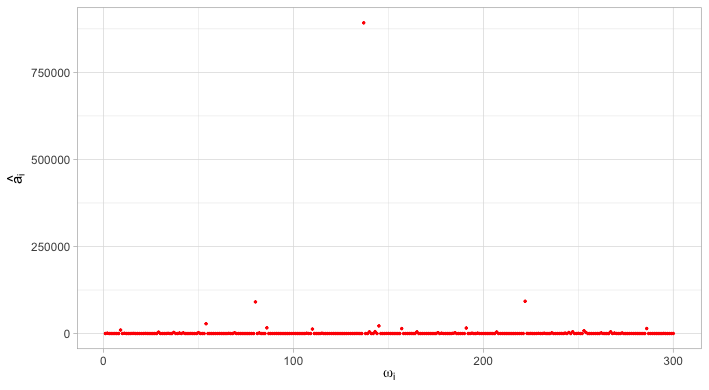}
  \caption{Half-Cauchy}
\end{subfigure}
\caption{Posterior mean estimates of the score parameters $a_i$'s for the generalised beta-CoRM with different hyperpriors.
}
\label{post_a_beta_corms}
\end{figure}

From these results we can immediately appreciate a highly-concentrated posterior distribution. This is certainly an interesting result since we can now compare this global behaviour against the individual score parameters of the generalised beta-CoRM models with different prior structures. To this end in Figure~\ref{post_a_beta_corms} we graphically present the posterior mean of the score parameters for the generalised beta-CoRM model with vague gamma prior, objective Lomax prior, Lomax prior and half-Cauchy type prior. From these plots it is compelling to see how with the vague gamma prior we obtain score parameters within the credible interval of $a$. In contrast, for the gamma-gamma prior cases we are able to see a more defined shrinkage effect ideal for the feature selection purposes of the generalised beta-CoRM.  However, it is clear that this effect is more pronounced for the objective Lomax and the half-Cauchy type priors, whereas for the Lomax prior we get a less variable behaviour for the score parameters. This is something that can be explained by the prior choices on $(\phi,\kappa)$ and the posterior behaviour on the ones that are not fixed as seen in Table~\ref{post_inference_phi} and Table~\ref{post_inference_kappa}.

\begin{table}[ht!]
    \centering
    \begin{tabular}{lccc}
    \hline
    Model/Prior & L. Interval & Median & U. Interval\\
    \hline
    Gen. beta-CoRM/Lomax & 37.7230  & 185.4517  & 727.0159 \\
    \hline
    \end{tabular}
    \caption{Median and 95$\%$ credible interval for $\phi$.}
    \label{post_inference_phi}
\end{table}

\begin{table}[ht!]
    \centering
    \begin{tabular}{lccc}
    \hline
    Model/Prior & L. Interval & Median & U. Interval\\
    \hline
    Gen. beta-CoRM/Lomax & 73.0926  & 370.5593  & 1432.916 \\
    Gen. beta-CoRM/Half-Cauchy & 1.4220  & 1.8380  & 2.3846 \\
    \hline
    \end{tabular}
    \caption{Median and 95$\%$ credible interval comparison for $\kappa$.}
    \label{post_inference_kappa}
\end{table}

One can immediately appreciate that the median values for $\phi$ and $\kappa$ are two orders of magnitude larger than the ones obtained/used for the rest of the models. However, from the scale-mixture gamma prior structure, one can understand this as a highly-concentrated distribution around the mean which is approximately $0.5$. That is why the beta-CoRM model with Lomax prior does not show such a heavy-tailed behaviour and extreme values for the score parameters as the ones obtained for the objective Lomax and the half-Cauchy type prior. Comparing these priors, one can appreciate that the half-Cauchy type prior yields larger score parameters $\hat{a}_i$. This can also be explained through the scale-mixture gamma structure, since for the half-Cauchy type prior we fixed $\theta=0.5$ and $\hat{\kappa}=1.8380$, whereas the objective Lomax had $\phi=\kappa=1$. Then the combination of a smaller $\theta$ and larger $\kappa$ yields local shrinkage parameters $\alpha_i$'s closer to zero, and hence, larger global score parameters $a_i$'s.

\subsection{Feature Selection and Predictive Performance}
Now that we have compared the posterior inference of the beta-CoRM models and the impact of the prior we can turn our attention to the feature selection procedure of the generalised beta-CoRM models. To this end we are mainly interested on the optimal number of features, the threshold at which they are found and of course, the impact on the predictive performance. Firstly, in Table~\ref{feats_thresh} we present the results on the first two characteristics mentioned above for the generalised beta-CoRM models. One can appreciate that the thresholds found get larger as we consider scale-mixture gamma priors with heavier tails. Clearly, with the half-Cauchy type prior having the largest threshold. A behaviour we have already seen in the global score parameters as explained in Section~\ref{post_inf_an}. 

Finally, to fully understand the impact of the prior in Table~\ref{beta_corms_perf} we present the predictive performance of the five beta-CoRM models using four widely used metrics in the literature when dealing with imbalanced sets \cite[\textit{see e.g.}][]{olson2008}, which are: accuracy (Acc.), precision (Prec.), recall (Rec.) and the F1-score ($F_1$). For this synthetic data set we are then able to see that the best results with respect the predictive performance of the model correspond to the generalised beta-CoRM with vague gamma and Lomax priors.

\begin{table}[ht!]
    \centering
    \begin{tabular}{lcc}
    \hline
    Model/Prior & Optimal Features & Threshold\\
    \hline
    Gen. beta-CoRM/Vague gamma  & 238  & 1.646140  \\
    Gen. beta-CoRM/Obj. Lomax  & 215  & 2.441105  \\
    Gen. beta-CoRM/Lomax & 280  & 2.804267   \\
    Gen. beta-CoRM/half-Cauchy & 234  & 122.388788   \\
    \hline
    \end{tabular}
    \caption{Optimal number of features and threshold for the generalised beta-CoRM model with different priors.}
    \label{feats_thresh}
\end{table}

\begin{table}[ht!]
    \centering
    \begin{tabular}{lcccc}
    \hline
   Model/Prior & Acc.($\%$)& Prec.($\%$)& Rec.($\%$)& $F_1$($\%$)  \\
    \hline
    Beta-CoRM/Vague gamma& 85.60  & 85.87  & 86.48  &85.84 \\
    Gen. beta-CoRM/Vague gamma   &87.60  & 88.42  & 89.57 &88.57\\
    Gen. beta-CoRM/Obj. Lomax  &86.40  & 86.06  & 87.61 &86.60\\
     Gen. beta-CoRM/Lomax   & 87.20  & 87.32  & 88.97 & 87.85\\
    Gen. beta-CoRM/Half-Cauchy  & 86.40  & 85.59  & 87.57 & 86.24\\
    \hline
    \end{tabular}
    \caption{Predictive performance of the five beta-CoRM models.}
    \label{beta_corms_perf}
\end{table}

\section{Malware Detection and Classification}
\label{Examples}

Malware is a computational term that is commonly used to describe any software specifically designed to disrupt, damage or gain access to a computer system. Traditionally, the use of antivirus software has been essential in order to detect malicious code and to keep the computer systems protected. Antivirus software usually makes use of the blacklisting method, where a new program is scanned in search of signatures of known malware and if found, the program is disabled and a warning is flagged. This approach is effective for detecting known threats; however, it has been proved to be less effective with new threats and with slight modifications made to the original code to avoid recognition \citep{McGraw}. A thorough review of statistical methods to deal with intrusion detection, including malware detection and classification, can be found in \cite{PGV_review}. 

In recent years, machine learning and statistical approaches have been used as an alternative to the blacklisting method and several approaches have been proposed in order to detect malicious code. One of such approaches is precisely by analysing the executable content of the malware through $n$-gram profiles \citep[see {\it e.g.}][]{Kolter2004, Pektas2011}. It is important to remark that malware detection is not the only task required when dealing with malicious software. In order to understand their infectious process, their potential threat level and therefore, how to be well-protected against these malicious software, there is a need to correctly identify the family to which they belong. The accurate classification may also speed-up the process of reverse-engineering to fix computer systems that were infected as well as for developing security patches to prevent more computers to become infected. Of course, this classification task can also be done through an $n$-gram profile analysis of the hexadecimal code.

Before introducing the data used it is important to remark that in computer science, the byte is the basic unit of information for storage and processing, and it is most commonly represented by a sequence of 8 binary digits or bits. Every instruction given to a computer can be broken down into sequences of bytes, which form the instruction's binary code. These binary sequences can be expressed in a more condensed form using the hexadecimal notation that is, each byte is represented as the combination of two elements of the set $\mathcal{T}=\{0,1,...,9,A,B,...,F\}$. For example, considering the hexadecimal representation of the code extract of a malicious software, given by $00 \hspace{.1cm} 00  \hspace{.1cm} 1C  \hspace{.1cm} 40  \hspace{.1cm} 2A  \hspace{.1cm} 28$, and $n=4$ we take all the possible sequences of 4 contiguous bytes to create the set of $4$-grams, that is $\{00 00 1C 40, 00 1C 40 2A, 1C 40 2A 28\}$. This data processing is then performed in the complete code for all the malicious and benign software (if applicable).

\subsection{Data}
\begin{figure}[ht]
\begin{subfigure}{.5\textwidth}
\centering
    \includegraphics[scale=.3]{ 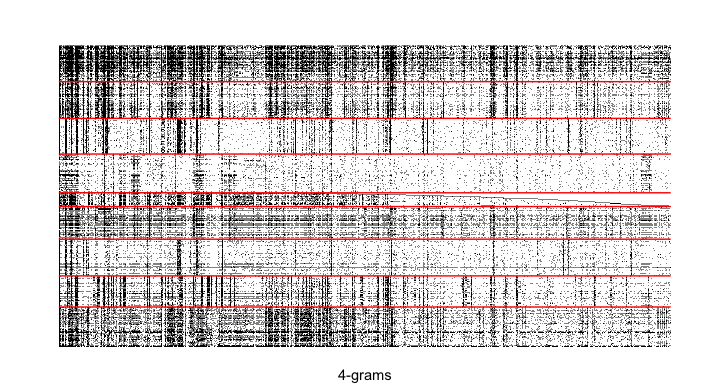}
    \caption{Training Data}
\end{subfigure}
\begin{subfigure}{.5\textwidth}
\centering
  \includegraphics[scale=.3]{ 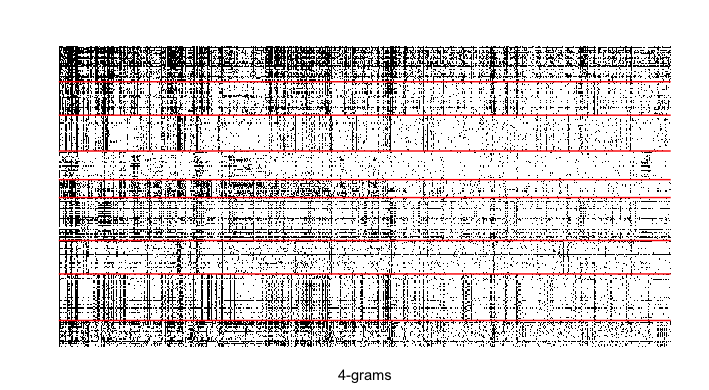}
  \caption{Test Data}
\end{subfigure}
\caption{Malware data set split into training set (left) and test set (right). For each plot the nine families are separated by the red horizontal lines. The black dots represent that a feature is present within the respective malware hexadecimal code.}
\label{Malware_data}
\end{figure}

The data used is a subset of the data released as part of the Microsoft Malware Classification Challenge \citep{MMCC} hosted at Kaggle in 2015. The data is composed of 842 malware representing a mix of nine different families. For each of these 842 malware we have the label representing the true family and the file with the hexadecimal representation of the binary code. These malware were further split into a training set composed of 590 observations and a test set of 292 elements. Taking into account this 590 malware the number of unique $4$-grams reached the order of 10 million that is why we decided to centre our attention on the $4$-grams that appeared at least once in each family yielding a unique set of 826 features. This drastic criterion allows us to practically eliminate the dependence of the 4-grams due to shared values in contiguous $n$-grams, since it can be seen that there are just 2524 (out of 112738) pairs of contiguous 4-grams. This justifies assuming independent (in a similar way to the bag-of-words assumption in text modelling), furthermore, this also allows us to consider $4$-grams that are likely to appear in new observations. Of course, other criteria could be used to reduce the number of initial features; however, this is something outside the scope of this paper. In Figure~\ref{Malware_data} both the training and the test set are graphically represented.  

From Figure~\ref{Malware_data} we are able to directly observe a clear and defined structure in some of the families, such as family three, where there are some predominant features that could help us discriminate new malware. However, it is also true that some of the families are not that well defined like families one, six and nine. This of course might be due to the way the initial set of features has been chosen, however, it also provides us with a nice motivation for the generalised beta-CoRM models since then we should be able to extract an optimal set of features to create a refined $n$-gram profiles analysis. 

\subsection{Posterior Inference Results}
In this section we present and discuss some interesting results of the four beta-CoRM models considered so far. In this case since the data set is almost 6.5 times bigger than the synthetic data we only considered 101,000 iterations for all the algorithms with a burn-in of 1000 and a thinning of 100 for an effective sample size of 1000. In this direction it is compelling to remark that all the algorithms roughly required 30 minutes to run contrary to the synthetic data where it was observed that the beta-CoRM and the generalised beta-CoRM with an objective Lomax prior and half-Cauchy prior required half the time to finish the posterior inference routines. Now, as for some of the results first we would like to compare the posterior samples of some common parameters. 

Firstly, in Table~\ref{post_c_malware} we present the median and the 95$\%$ credible interval of the concentration parameter $c$ for all the models considered so far. We can then appreciate a slight difference between the median and the intervals of the beta-CoRM models with vague gamma priors and the generalised version with gamma-gamma priors.

\begin{table}[ht]
    \centering
    \begin{tabular}{lccc}
    \hline
    Model/Prior & L. Interval & Median & U. Interval\\
    \hline
    Beta-CoRM/Vague gamma & 2.0161 & 2.2292  & 2.4829 \\
    Gen. beta-CoRM/Vague gamma & 2.0273  & 2.2252  & 2.4512 \\
    Gen. beta-CoRM/Obj. Lomax  & 1.9517  & 2.1691  & 2.4140 \\
    Gen. beta-CoRM/Lomax & 1.9814  & 2.1873  & 2.4285 \\
    Gen. beta-CoRM/Half-Cauchy & 1.9485  & 2.1587  & 2.3963 \\
    \hline
    \end{tabular}
    \caption{Median and 95$\%$ credible interval comparison for the concentration parameter.}
    \label{post_c_malware}
\end{table}

Now we centre our attention on the score parameter for the beta-CoRM and the score parameters for the generalised versions, since as detailed throughout the paper these parameters play a vital role in the feature selection step. In Table~\ref{post_a_malware_beta_CoRM} we first present the median and credible interval of the score parameter for the beta-CoRM model, from which we can notice that the posterior distribution exhibits a low variance.  

\begin{table}[ht]
    \centering
    \begin{tabular}{lccc}
    \hline
    Model & Lower Interval & Median & Upper Interval\\
    \hline
    Beta-CoRM & 0.7407  & 0.7599  & 0.7792 \\
    \hline
    \end{tabular}
    \caption{Median and 95$\%$ credible interval comparison for the score parameter}
    \label{post_a_malware_beta_CoRM}
\end{table}

\begin{figure}[ht]
\begin{subfigure}{.5\textwidth}
\centering
  \includegraphics[scale=.3]{ 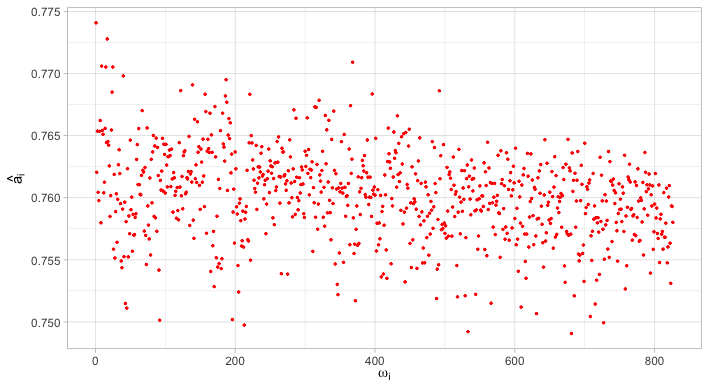}
  \caption{Vague gamma}
\end{subfigure}
\begin{subfigure}{.5\textwidth}
\centering
  \includegraphics[scale=.3]{ 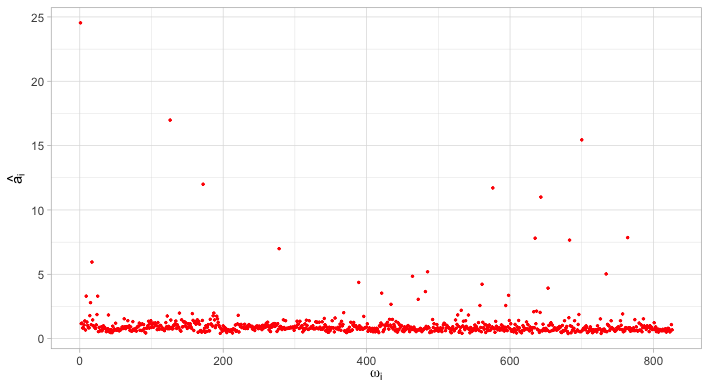}
  \caption{Objective Lomax}
\end{subfigure}
\begin{subfigure}{.5\textwidth}
\centering
  \includegraphics[scale=.3]{ 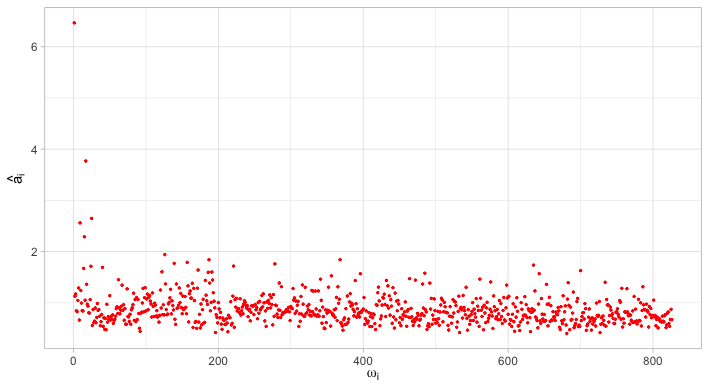}
  \caption{Lomax}
\end{subfigure}
    \begin{subfigure}{.5\textwidth}
\centering
  \includegraphics[scale=.3]{ 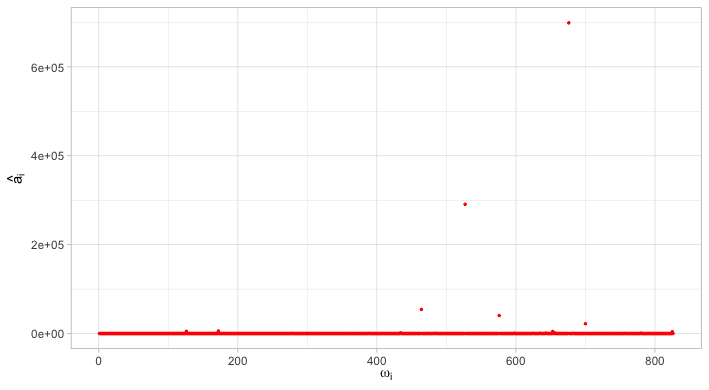}
  \caption{Half-Cauchy}
\end{subfigure}

\caption{Posterior mean estimates of the score parameters $a_i$'s for the generalised beta-CoRM with different hyperpriors.
}
\label{post_a_malware}
\end{figure}

Lastly, in Figure~\ref{post_a_malware} we present the posterior mean estimates of the score parameters $a_i$'s for the generalised beta-CoRM under the four prior structures considered so far. We can then again notice how the generalised beta-CoRM model with vague gamma priors yields score parameters, $a_i's$, within the credible interval of the score parameter $a$ of the beta-CoRM. Then and just as expected, the gamma-gamma prior exhibits a shrinkage behaviour which certainly helps the motivation behind the feature selection process. In particular, we can again appreciate how the effect of having quite isolated scores is more predominant for the objective Lomax and half-Cauchy type prior, whereas for the Lomax the variability is more controlled. 

\subsection{n-gram Profiles Analysis and Feature Selection}

\begin{figure}[ht!]
\begin{center}
\begin{subfigure}{.5\textwidth}
\centering
    \includegraphics[scale=.3]{ 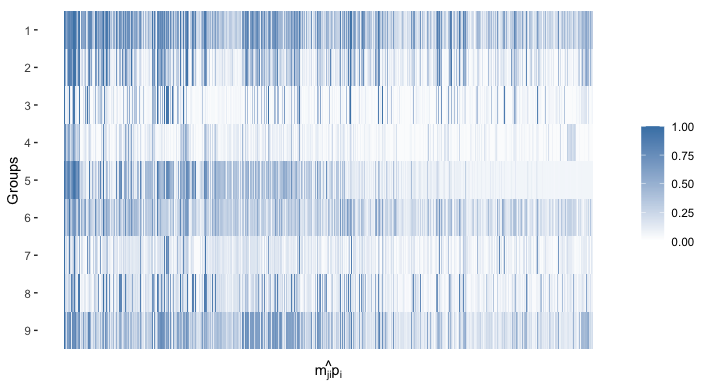}
    \caption{Beta-CoRM}
\end{subfigure}%
\end{center}
\begin{subfigure}{.5\textwidth}
\centering
  \includegraphics[scale=.3]{ 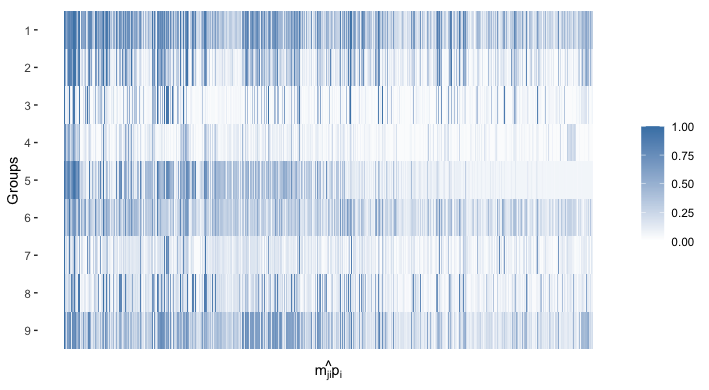}
      \caption{Vague gamma}
\end{subfigure}
\begin{subfigure}{.5\textwidth}
\centering
  \includegraphics[scale=.3]{ 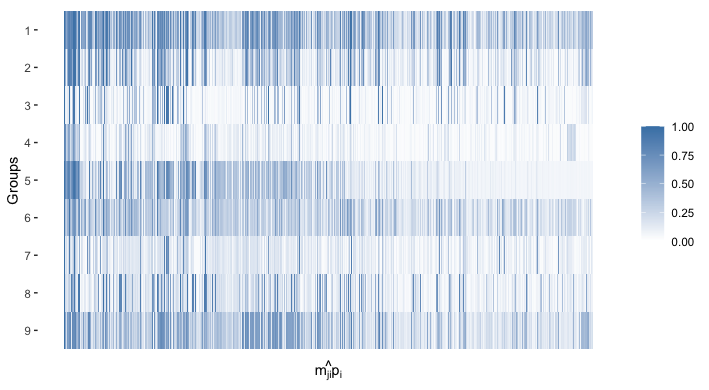}
  \caption{Objective Lomax}
\end{subfigure}
\\
\begin{subfigure}{.5\textwidth}
\centering
  \includegraphics[scale=.3]{ 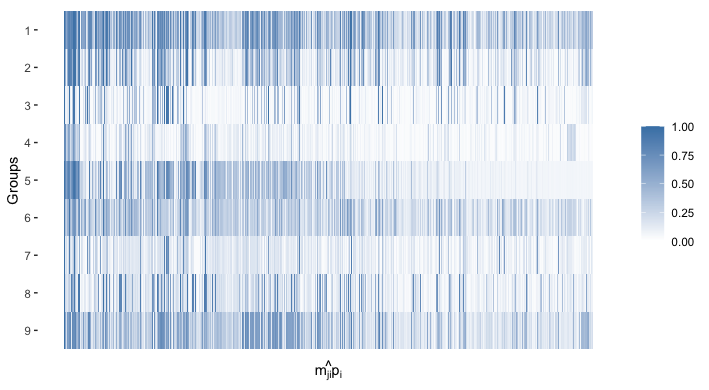}
  \caption{Lomax}
\end{subfigure}
    \begin{subfigure}{.5\textwidth}
\centering
  \includegraphics[scale=.3]{ 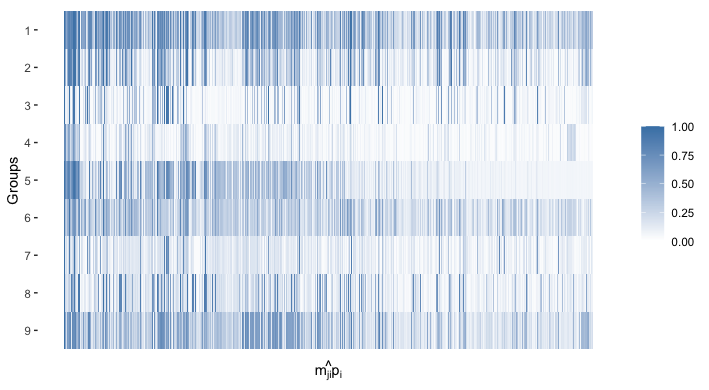}
    \caption{Half-Cauchy}

\end{subfigure}
\caption{$4$-gram profiles represented by the posterior predictive probabilities for the beta-CoRM and the gen. beta-CoRM with different hyperpriors.
}
\label{malware_profiles}
\end{figure}

Now that we have a better sense on the inference, we can proceed to the analysis of the $n$-gram profiles produced by the beta-CoRM models. First, in Figure~\ref{malware_profiles} we display the graphical representation of the posterior predictive probabilities for each of the 826 $4$-grams across the nine families of malware. It is immediate to notice that the five $4$-gram profiles are quite similar and visually it is difficult to see any meaningful differences among each other. To have a better sense of this we obtain the pair-wise Euclidean distance of the profiles and present the results in Table~\ref{diff_profiles}. We can appreciate that beta-CoRM models with vague gamma priors yield the most similar profiles followed by the model with objective Lomax and half-Cauchy type prior.

\begin{table}[ht!]
    \centering
    \begin{tabular}{cccccc}
    \hline
    Profile &  1 & 2 & 3 &  4 & 5\\
    \hline
     1 & 0 & 0.1704431  & 0.4213533  & 0.3702292&0.4369723\\
      2 & 0.1704431  & 0  & 0.4160238  & 0.3622737&0.4304231\\
      3 & 0.4213533 & 0.4160238  & 0  & 0.1938622&0.1721644\\
      4 & 0.3702292 & 0.3622737  & 0.1938622  & 0&0.2027637\\
      5 & 0.4369723 & 0.4304231  & 0.1721644  &0.2027637 &0\\
     \hline
    \end{tabular}
    \caption{Euclidean distance of the $n$-gram profiles of the beta CoRM (Profile 1) and the gen. beta-CoRM models with vague gamma (Profile 2), objective  Lomax (Profile 3), Lomax  (Profile 4) and half-Cauchy (Profile 5) priors.}
    \label{diff_profiles}
\end{table}

Of course, it is important to remember that the main advantage of the generalised models is that they allow us to find an optimal number of features to reduce the uncertainty in the data and which can be further used to improve the predictive performance of the model. To this end and contrary to the synthetic data we obtain the optimum threshold and the optimal features by using the training set as our validation set as well. With this in mind, we present in Table~\ref{accuracy_genbetaCoRM_trainig} the results on the training set including the maximum accuracy achieved, the threshold used for the generalised beta-CoRM models  and the number of features required to achieve the maximum accuracy. Clearly, with the generalised versions we are able to better explain the data with the objective Lomax prior being the best one. Furthermore, we can again appreciate the similarities between the models with vague gamma priors and the ones having the gamma-gamma prior structure. 
 
\begin{table}[ht]
    \centering
    \begin{tabular}{lccc}
    \hline
    Model/Prior &  Accuracy ($\%$) & Threshold & Features\\
    \hline
     Beta-CoRM/Vague Gamma & 80.51 & - & 826\\
     Gen. beta-CoRM/Vague Gamma & 85.59 & 0.756499  & 116\\
     Gen. beta-CoRM/Obj. Lomax & 85.93 & 0.609369  & 124 \\
      Gen. beta-CoRM/Lomax & 84.75 & 0.615221 & 129  \\
      Gen. beta-CoRM/Half-Cauchy & 85.08 & 0.592240 & 124  \\
     \hline
    \end{tabular}
    \caption{Prediction results on the training set by the generalised beta-CoRM models with vague gamma, objective Lomax, Lomax and half-Cauchy type priors respectively.}
    \label{accuracy_genbetaCoRM_trainig}
\end{table}

\begin{figure}[ht!]
\begin{subfigure}{.5\textwidth}
\centering
    \includegraphics[scale=.3]{ 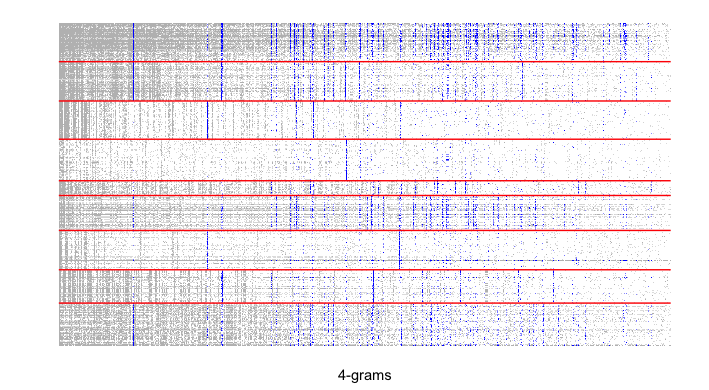}
    \caption{Vague gamma}
\end{subfigure}%
\begin{subfigure}{.5\textwidth}
\centering
  \includegraphics[scale=.3]{ 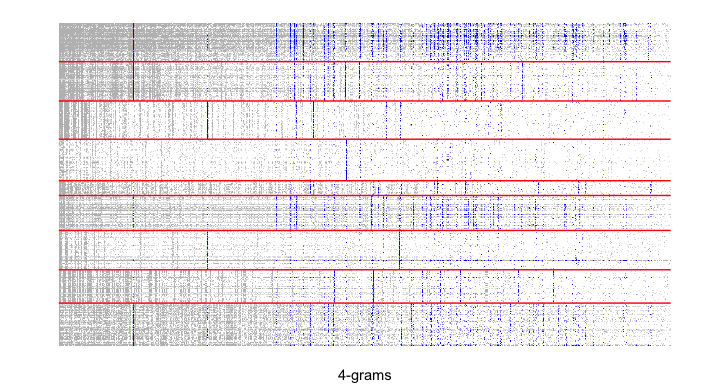}
      \caption{Objective Lomax}
\end{subfigure}
\begin{subfigure}{.5\textwidth}
\centering
  \includegraphics[scale=.3]{ 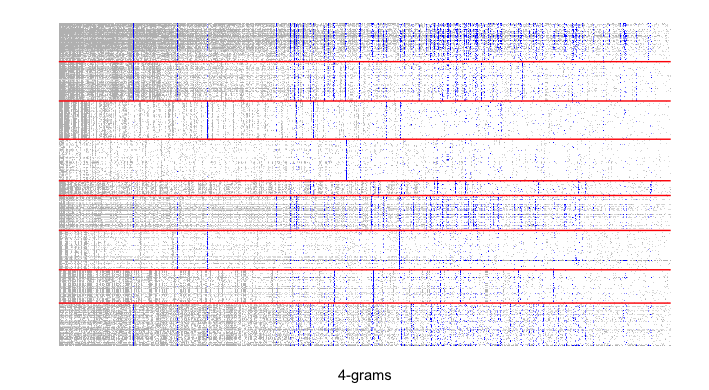}
  \caption{Lomax}
\end{subfigure}
\begin{subfigure}{.5\textwidth}
\centering
  \includegraphics[scale=.3]{ 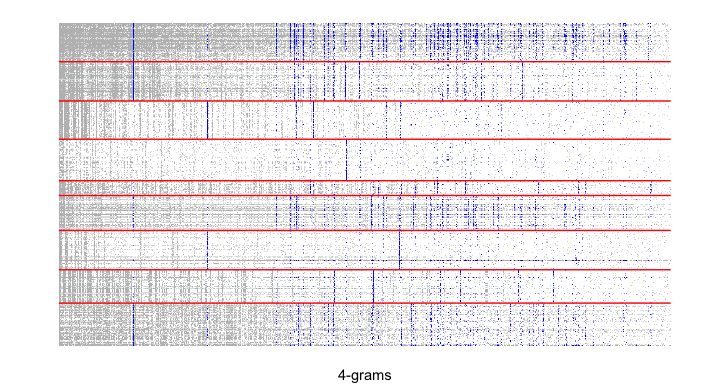}
  \caption{Half-Cauchy}
\end{subfigure}
\caption{Restricted data on the optimal features (blue) for generalised beta-CoRM model with different hyperpriors.
}
\label{restricted_data}
\end{figure}

With respect the feature selection it is further interesting to notice that the four generalised models coincide in 98 features, and this number increases to 119 common features if we just consider the generalised beta-CoRM models with gamma-gamma priors. To fully understand the process of feature selection on this noisy data set we present in Figure~\ref{restricted_data} the data restricted to the optimal figures for the four generalised beta-CoRM models. Then we are able to see better defined groups with the exception of the first, sixth and last family. Nevertheless the models are still able to identify some influencing features in the groups, which we can exploit to improve the predictive performance of the original model on the test set as seen in Table~\ref{overall_malware_metrics}.

\begin{table}[ht!]
    \centering
    \begin{tabular}{lcccc}
    \hline
    Model/Prior &  Acc.($\%$) & Prec.($\%$) & Rec.($\%$) & F$_1$($\%$)\\
    \hline
     beta-CoRM/Vague gamma & 80.16 & 82.23  & 81.22 & 80.44\\
     G. beta-CoRM/Vague gamma & 83.73 & 84.82  & 84.86&84.00\\
      G. beta-CoRM/Obj. Lomax & 84.13 & 84.85 & 84.97 &84.11\\
      G. beta-CoRM/Lomax & 83.33 & 84.14 & 84.35&83.36  \\
      G. beta-CoRM/Half-Cauchy & 83.73 & 84.84 & 85.09 &  84.07 \\
     \hline
    \end{tabular}
    \caption{Predictive performance of the five beta-CoRM models.}
    \label{overall_malware_metrics}
\end{table}

Finally, from Table~\ref{overall_malware_metrics} we can further appreciate that the beta-CoRM model with the objective Lomax prior achieves the best results overall. So that we can proceed to compare it against some well-known supervised learning models. For the latter it is important to remark that we have chosen binary classifiers that can be easily extended to a multi-class setting and that have already been used for detecting and classifying malware. That is why the algorithms considered are: \textit{multinomial logistic regression} \citep[see {\it e.g.}][]{hosmer2013applied}; and \textit{decision trees} \citep[see {\it e.g.}][]{mitchell1997machine} with their \textit{extreme gradient} \citep{XGBOOST} boosted version. For their implementation we respectively used the R packages: \textit{nnet} \citep{nnet}, \textit{tree} \citep{tree}, and \textit{xgboost} \citep{xgboost1}. The results can be seen in Table~\ref{overall_ml_malware_metrics}. The generalized beta-CoRM model outperforms decision trees and has similar performance to multinomial logistic regression. However, the method is outperformed by  XGBoost.

\begin{table}[ht!]
    \centering
    \begin{tabular}{lcccc}
    \hline
    Model &  Acc.($\%$) & Prec.($\%$) & Rec.($\%$) & F$_1$($\%$)\\
    \hline
     Decision tree & 80.16 & 80.71  & 81.23 & 80.22\\
     Multinomial & 87.7 & 88.86  & 87.7 &87.84\\
      XGBoost  & 93.25 & 93.39 & 93.27 &93.12\\
     \hline
    \end{tabular}
    \caption{Predictive performance of decision tree, the multinomial logistic regression and XGBoost.}
    \label{overall_ml_malware_metrics}
\end{table}

\section{Conclusions}
\label{conclusion}

$n$-gram profiles are an important tool that have been used  to discover unique properties of the observations and the classes to which they belong. Depending on the application the modelling of $n$-gram profiles can either exploit their presence or absence or their frequencies. In this paper, we have centred our attention on the former since this approach has been used for cybersecurity applications that clearly represent a challenging and interesting opportunity for Bayesian statistics due to the growing demand for robust algorithms and methods capable of defending computer networks. Of course, as we have established throughout the paper, the use of $n$-grams can yield a large number of features to be considered which is infeasible from a practical point of view. To address this, we have designed a flexible model that allows us to consider a suitable and straightforward generalisation to perform a feature selection procedure to find an optimal subset of features.

Furthermore, we have considered a deep prior sensitivity analysis for which we have discovered how certain priors exhibit more clearly the shrinkage effect required for the feature selection. Although, it is clear that after a long time running the posterior simulations we have reached quite consistent results regardless of the prior chosen. Of course, it is important to remark that other non-conjugate shrinkage priors could be used; however, the computational cost might increase. This is something that we need to always keep in mind whenever dealing with applications like cyber security where fast and scalable algorithms are required. Finally, and from a cyber security point of view there is definitely more interesting research and applications where we could exploit the beta-CoRM models. For example, $n$-gram profiles have also been used to detect masquerade attacks, where synthetic identities are used to fraudulently access networks. Unsupervised learning models are developed which distinguish normal and  malicious activity. We believe that the beta-CoRM models could be used as well in this kind of applications, where we can flexibly model a ``normal'' distribution over $n$-grams. However, this might require a slightly different approach since in this paper we are assuming we have access to both benign and malicious observations. Both this application and the malware application considered in this paper could use the beta-CORM model to take into account the effect of context, such as the current operation of a device, expressed as a discrete variable. 
More perturbation coefficients would be introduced to model
the relationship between different contexts (and potentially different malware families).

\section*{Acknowledgements}
\label{acknowledgemets}
We would like to acknowledge the University of Kent where most of this work was completed as part of José Perusquía's PhD project funded by a GTA Scholarship. We would also like to thank the Editor, the Associate Editor and the anonymous reviewers for their thoughtful suggestions and comments to improving the paper.
\appendix
\section{Proofs}
\label{proofs}
\subsection*{Proof of Proposition 1}

Proving the properties established in Proposition 1, only requires to remark that the variables $p_i$, are beta distributed with parameters $(cq_i,c(1-q_i))$. Therefore $\Esp(p_i)=q_i$, and using the monotone convergence theorem we get 
\begin{equation*}
\Esp(B)=\Esp\left(\sum_{i=1}^{\infty}p_i\delta_{\omega_i}\right)=\sum_{i=1}^{\infty}\Esp(p_i)\delta_{\omega_i}=\sum_{i=1}^{\infty}q_i\delta_{\omega_i}=B_0.
\end{equation*}
Following the same monotone convergence reasoning and using the fact that $\Esp(p_i^2)=\frac{q_i(1-q_i)}{c+1}+q_i^2$ it can be shown that
\begin{equation*}
\Esp(B^2)=B_0^2+\frac{1}{c+1}\sum_{i=1}^{\infty}q_i(1-q_i).
\end{equation*}
From which the variance can be obtained directly. 

\subsection*{Proof of Proposition 2}

The first property is straightforward since it follows the same reasoning as in Proposition 1. Now, for a fixed atom $\omega_i$  we have that the variance is given by,

\begin{eqnarray*}
\Var(B_j(d\omega_i))&=&\Esp(B_j^2(d\omega_i))-\Esp(B_jd(\omega_i))^2=\Esp(m_{ji}^2p_i^2)-\Esp(m_{ji}p_i)^2\\
&=&\left(\frac{a}{(a+1)^2(a+2)}+\frac{a^2}{(a+1)^2}\right)\left(\frac{q_i(1-q_i)}{c+1}+q_i^2\right)-\left(\frac{a}{a+1}\right)^2q_i^2\\
&=&\left(\frac{a}{(a+1)^2(a+2)}+\frac{a^2}{(a+1)^2}\right)\left(\frac{q_i(1-q_i)}{c+1}\right)+\left(\frac{a}{(a+1)^2(a+2)}\right)q_i^2\\
&=&\left(\frac{a}{a+2}\right)\left(\frac{q_i(1-q_i)}{c+1}\right)+\left(\frac{a}{(a+1)^2(a+2)}\right)q_i^2\\
&=&\left(\frac{aq_i}{a+2}\right)\left(\frac{(1-q_i)(a+1)^2+q_i(c+1)}{(c+1)(a+1)^2}\right).
\end{eqnarray*}
$\Var(B_j(d\omega_i))=\Var(B_k(d\omega_i))$. 

\subsection*{Proof of Proposition 3}
From Proposition 2, it can be clearly appreciated that for a fixed feature, the variance is the same across families. This will be useful in order to obtain the correlation. But first, for the covariance between $B_j(d\omega_i)$ and $B_k(d\omega_i)$ we have from Proposition 2 that,
\begin{equation*}
\Esp(B_j(d\omega_i))=\left(\frac{a}{a+1}\right)B_0(d\omega_i)=\left(\frac{a}{a+1}\right)q_i=\Esp(B_k(d\omega_i))
\end{equation*}
and
\begin{equation*}
\Esp(B_j(d\omega_i)B_k(d\omega_i))=\Esp(m_{ji}m_{ki}p_i^2)=\left(\frac{a}{a+1}\right)^2\left(\frac{q_i(1-q_i)}{c+1}+q_i^2\right),
\end{equation*}
therefore,
\begin{eqnarray*}
\Cov(B_j(d\omega_i),B_k(d\omega_i))&=&\Esp(B_j(d\omega_i)B_k(d\omega_i))-\Esp(B_j(d\omega_i))\Esp(B_k(d\omega_i))\\
&=&\left(\frac{a}{a+1}\right)^2\left(\frac{q_i(1-q_i)}{c+1}+q_i^2\right)-\left(\frac{a}{a+1}\right)^2q_i^2\\
&=&\left(\frac{a}{a+1}\right)^2\left(\frac{q_i(1-q_i)}{c+1}\right).
\end{eqnarray*}
Hence, the correlation is given by
\begin{eqnarray*}
\Corr(B_j(d\omega_i),B_k(d\omega_i))&=&\frac{\Cov(B_j(d\omega_i),B_k(d\omega_i))}{\Var(B_j(d\omega_i))}\\
&=&\left(\frac{a}{a+1}\right)^2\left(\frac{q_i(1-q_i)}{c+1}\right)\left(\frac{a+2}{aq_i}\right)\left(\frac{(c+1)(a+1)^2}{(1-q_i)(a+1)^2+q_i(c+1)}\right)\\
&=&\frac{a(a+2)(1-q_i)}{(1-q_i)(a+1)^2+q_i(c+1)}.
\end{eqnarray*}

\subsection*{Proof of Lemma 1.}
Let us consider first the augmented model. In this case it is straightforward to see that conditioned on $y_{kji}=0$ we have that $x_{kji}\stackrel{\text{a.s.}}{=}0$ and conditioned on $y_{kji}=1$ we have that $x_{kji}\sim\text{Ber}(p_i)$. With this in mind, the augmented likelihood is
\begin{eqnarray*}
\prod_{j=1}^d\prod_{k=1}^{n_j}\left(\delta^{x_{kji}}_{0}\right)^{(1-y_{kji})}\left(p_i^{x_{kji}}(1-p_i)^{(1-x_{kji})}\right)^{y_{kji}}
\end{eqnarray*}
and the posterior distribution is proportional with respect to the latent variables to
\begin{eqnarray*}
\prod_{j=1}^d\prod_{k=1}^{n_j}\left(\delta^{x_{kji}}_{0}\right)^{(1-y_{kji})}\left(p_i^{x_{kji}}(1-p_i)^{(1-x_{kji})}\right)^{y_{kji}}m_{ji}^{y_{kji}}(1-m_{ji})^{(1-y_{kji})}.
\end{eqnarray*}
Integrating out the latent variables yields
\begin{eqnarray}\label{post_slice}
\prod_{j=1}^d\prod_{k=1}^{n_j}\delta^{x_{kji}}_{0}(1-m_{ji})+m_{ji}p_i^{x_{kji}}(1-p_i)^{(1-x_{kji})}.
\end{eqnarray}
Therefore, we can appreciate that the marginal posterior is the product of the mixture of a degenerate distribution and a Bernoulli distribution with corresponding weights $(1-m_{ji})$ and $m_{ji}$. This expression at first sight does not resemble the posterior distribution for the beta-CoRM model. However, it is sufficient to notice that from \eqref{post_slice} we obtain
\begin{align*}
\delta^{x_{kji}}_{0}(1-m_{ji})+m_{ji}p_i^{x_{kji}}(1-p_i)^{(1-x_{kji})}=\begin{cases}
1-m_{ji}p_i&\text{if } x_{kji}=0\\
m_{ji}p_i&\text{if } x_{kji}=1.
\end{cases}
\end{align*}
Hence, we recover the original posterior distribution.
\bibliographystyle{hapalike}
\bibliography{References.bib}

\end{document}